\documentclass{article}
\usepackage[T1]{fontenc} % add special characters (e.g., umlaute)
\usepackage[utf8]{inputenc} % set utf-8 as default input encoding
\usepackage{ismir,amsmath,cite,url}
\usepackage{graphicx}
\usepackage{color}
\usepackage{hyperref}

\usepackage{acronym} 
\acrodef{dtw}[DTW]{Dynamic Time Warping}
\acrodef{hmm}[HMM]{Hidden Markov Model}
\acrodef{nmf}[NMF]{Non-Negative Matrix Factorization}
\acrodef{olda}[OLDA]{Ordinal Linear Discriminant Analysis}
\acrodef{msd}[MSD]{Music Structure Discovery}
\acrodef{msa}[MSA]{Music Structure Analysis}
\acrodef{ssm}[SSM]{Self-Similarity-Matrix}
\acrodef{mir}[MIR]{Music Information Retrieval}
\acrodef{ssl}[SSL]{Self-Supervised-Learning}
\acrodef{sa}[SA]{Self-Attention}
\acrodef{tl}[TL]{Triplet Loss}
\acrodef{fc}[FC]{Fully-Connected}
\acrodef{cqt}[CQT]{Constant-Q-Transform}
\acrodef{fc}[FC]{Fully-Connected}
\acrodef{bce}[BCE]{Binary-Cross-Entropy}
\acrodef{dnn}[DNN]{Deep Neural Network}
\acrodef{cnn}[ConvNet]{Convolutional Networks}
\acrodef{msd}[MSD]{Music Structure Discovery}
\acrodef{lms}[LMS]{Log-Mel-Spectrogram}
\acrodef{pcp}[PCP]{Pich-Class-Profile}
\acrodef{lsd}[LSD]{Laplacian Structural Decompositon}

\acrodef{hc}[HC]{hand-crafted}
\acrodef{dl}[DL]{deep learning}
\acrodef{af}[AF]{audio features}
\acrodef{ds}[DS]{detection system}

\renewcommand{\L}[0]{\mathcal{L}}
\newcommand{\X}[0]{\mathbf{X}}
\renewcommand{\S}[0]{\mathbf{S}_{ij}}
\newcommand{\hS}[0]{\mathbf{\hat{S}}_{ij}^{\theta}}
\newcommand{\n}[0]{\mathbf{n}_{i}}
\newcommand{\hn}[0]{\mathbf{\hat{n}}_{i}^{\theta}}
\newcommand{\e}[0]{\mathbf{e}^{\theta}}

\newcommand{\f}[0]{f^{\theta}}
\newcommand{\ddd}[2]{\frac{\partial #1}{\partial #2}}
\newcommand{\K}[0]{\mathbf{K}^{\theta}}

\newcommand{\HRt}[0]{\texttt{HR3F}}
\newcommand{\HRd}[0]{\texttt{HR0.5F}}

\usepackage{booktabs}

\usepackage{titlesec}
\title{Self-Similarity-Based and Novelty-based loss for music structure analysis} 

\oneauthor
{Geoffroy Peeters}
{LTCI, Télécom-Paris, Institut Polytechnique de Paris, France}
\def\authorname{G. Peeters}

\begin{document}

\maketitle

% ---------------------------------
\begin{abstract}
	\ac{msa} is the task aiming at identifying musical segments that compose a music track and possibly label them based on their similarity.
	In this paper we propose a supervised approach for the task of music boundary detection.
	In our approach we simultaneously learn features and convolution kernels.
	For this we jointly optimize 
	- a loss based on the \ac{ssm} obtained with the learned features, denoted by SSM-loss, and
	- a loss based on the novelty score obtained applying the learned kernels to the estimated SSM, denoted by novelty-loss.
	We also demonstrate that relative feature learning, through self-attention, is beneficial for the task of MSA.
	Finally, we compare the performances of our approach to previously proposed approaches on the standard RWC-Pop, and various subsets of SALAMI.
\end{abstract}

% ---------------------------------
\section{Introduction}\label{sec:introduction}

\acf{msa} is the task aiming at identifying musical segments that compose a music track (a.k.a. segment boundary estimation) and possibly label them based on their similarity (a.k.a. segment labeling).
We deal here with \ac{msa} from audio.
\ac{msa} is one of the oldest task in \acl{mir}\footnote{Foote's paper~\cite{FooteACM99Visualizing} on \ac{ssm} was published in 1999.} but still one of the most challenging.
This is due to the difficulty to exactly define what music structure is and hence be able to create annotated datasets to measure progress or train systems.
People agree that the structure can be considered from multiple viewpoints\footnote{musical role, acoustic similarity, instrument role, perceptual tests} \cite{Peeters09LSASStructure}\cite{smith_design_2011}, is hierarchical~\cite{mcfee_hier} and is partly subjective~\cite{Bruderer2006ISMIRPerceptionStructure}.
Probably because of this complexity, the number of contributions in \ac{msa} has remained low despite its large number of applications: audio summarization \cite{peeters_toward_2002}, interactive browsing \cite{Peeters2008ISMIRMCIpa, Peeters2012IsmirMsse, goto2011songle}, musical analysis~\cite{Mueller2021IsmirScapePlot}, tools for researcher (to help chord recognition~\cite{Mauch2009IsmirChordStructure}, source separation~\cite{Rafii2013TASLPRepeat} or downbeat estimation~\cite{downbeat_fuentes}).

% --- Homogeneity, Repetition
To solve the two \ac{msa} tasks (boundary detection and segment labeling), three assumptions~\cite{Paulus2010ISMIRSTAR} are commonly used:
(1) \textit{novelty} (we assume that segments are defined by large ---novel--- changes of the musical content over time), 
(2) \textit{homogeneity} (the musical content is homogeneous within a given segment) and 
(3) \textit{repetition} (the musical content ---homogeneous or not--- can be repeated over time).
This has been extended by~\cite{Sargent2011IsmirStrcuture} to a fourth \textit{regularity} assumption (the segment's durations are regular over time).
Combining those allows to construct \ac{msa} systems.

% ----------------------------------------------------------
% ----------------------------------------------------------
\subsection{Related works}
\label{sec:related}

Over time, a large palette of approaches has been proposed for \ac{msa}.
We only review the ones related to our work and refer the reader to Nieto et al.~\cite{nieto2020audio} for a good overview.
We consider three periods according to 
the nature of the \acl{af} --\ac{hc} or learned by \ac{dl} --, and 
the nature of the \acl{ds} which uses the \acl{af} -- \ac{hc} or trained by \ac{dl} --.

\textbf{First period: \ac{hc} \acl{ds} applied to \ac{hc} \acl{af}.}
In these systems \ac{hc} \acl{af} (such as MFCC or Chroma) were given as input to \ac{hc} \acl{ds} (such as the checkerboard kernel, novelty-score~\cite{foote_automatic_2000}), 
unsupervised training (such as HMM~\cite{peeters_toward_2002}, NMF~\cite{kaiser_music_2010}),
supervised (such as OLDA~\cite{McFee2014ICASSPStructure}) or
pattern matching algorithms (such as DTW~\cite{Muller2013TASLPStructure} or variants \cite{serra_unsupervised_2014}) .

\textbf{Second period: \ac{dl} \acl{ds} applied to \ac{hc} \acl{af}.}
Over time, more and larger annotated datasets for \ac{msa} have been developed; which concomitantly with the development of \ac{dl} has allowed to re-formulate the \ac{msa} task in terms of supervised learning.
The \acl{ds} developed here mainly target the task of boundary detection.
For example, \cite{ullrich_boundary_2014} \cite{grill_music_2015} \cite{cohen-hadria_music_2017} propose to train in a supervised way a \ac{cnn} $\hat{y}=f^{\theta}(\X)$  to estimate if the center of a patch of \ac{hc} \acl{af} $\X$ is a boundary ($y$=1).
Various \ac{hc} \acl{af} (or combinations of) are used here: \acl{lms}, \acl{pcp} through \ac{ssm} expressed in (time,time) or (time,lag).

\textbf{Third period:  \ac{hc} \acl{ds} applied to \ac{dl} \acl{af}.
}
To deal with the endless debate about the choice of \ac{hc} \acl{af}, McCallum et al.~\cite{mccallum_unsupervised_2019} propose to learn them. 
For this, they train an encoder $\f$ by minimizing a \ac{tl} \cite{schroff_facenet_2015} between patches of beat-synchronous \ac{cqt}.
For the \ac{tl}, they propose a \ac{ssl} paradigm\footnote{which does not require any annotated segments and labels} to define the anchor $A$ patch, positive $P$ patch and negative $N$ patch.
Using the homogeneity assumption, neighboring times are supposed to be more similar to each other (therefore used to define $A$ and $P$) than to distant ones (used to define $N$).
For training they use a very large unlabeled dataset of 28345 songs.
This method however does not consider the repetition assumption\footnote{$N$ could potentially be in a segment which is a repetition of the segment containing $A$}.

Wang et al.~\cite{Wang2021ISMIRSupervised} revised McCallum approach  in a supervised setting. 
In this, the patches $P$ (resp. $N$) are now explicitly chosen so as to have the same (resp. different) annotated segment label than the patches $A$.
This supervised method now consider both the homogeneity and repetition assumption.
In another work  \cite{Wang2022ICASSPChorus}, they propose a spectral-temporal Transformer-based model (SpecTNT) trained with a connectionist temporal localization (CTL) loss to jointly estimate music segments ad their labels.

McCallum approach has also been extended by Buisson et al.~\cite{Buisson2022IsmirStructure} to take benefit from the hierarchy of structure in music. 
They show that the obtained deep embeddings can improve segmentation at various levels of granularity.

Rather than learning features for \ac{msa}, Salamon et al.~\cite{Salamon2021ISMIRDeepEmbedding} proposed to re-use pretrained ones.
Those are obtained using encoders previously trained on different tasks (Few-Shot Learning sound event and music auto-tagging).
Those are then used as input to a \acl{lsd} algorithm for \ac{msa}.

% ---------------------------------
% ---------------------------------
\subsection{Proposal and paper organization}
\label{sec:proposal}

\begin{figure*}[ht]
	\centerline{\includegraphics[trim=0cm 0cm 0cm 0cm,width=1.1\textwidth]{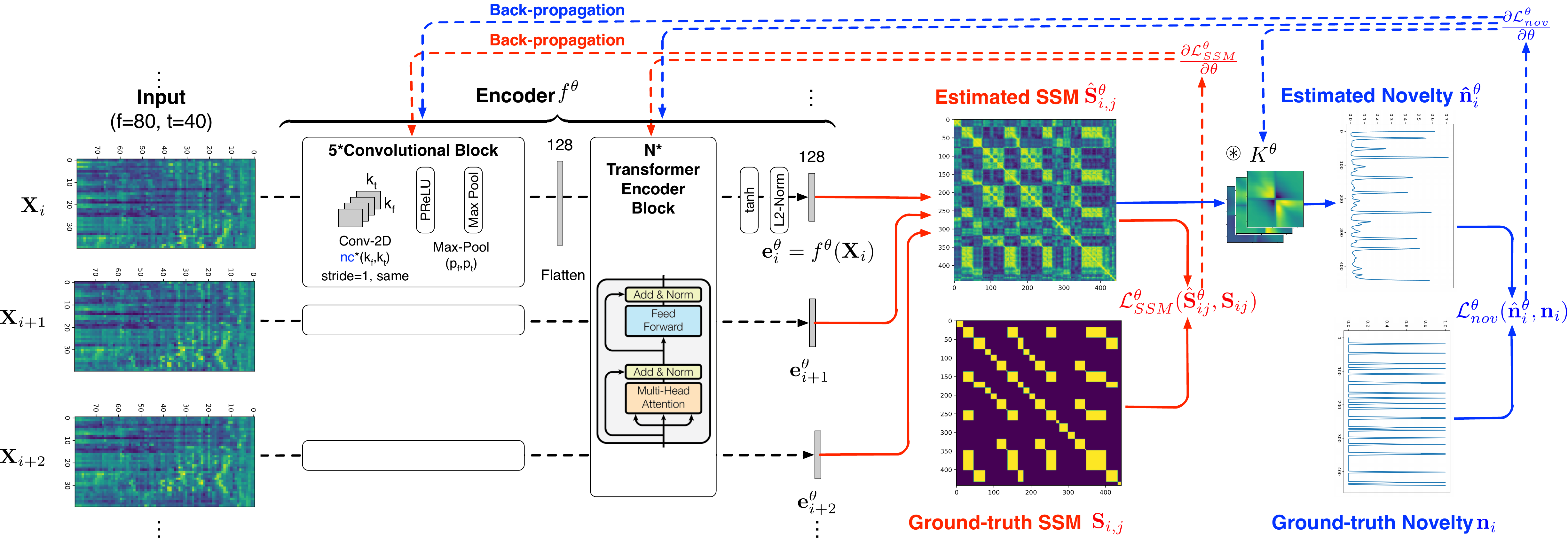}}
	\caption{Proposed architecture and training paradigm minimizing a \ac{ssm} loss $\L^{\theta}_{SSM}$ and a novety loss $\L^{\theta}_{nov}$ . 
	%Real illustrations from track 1 from RWC-Pop which is part of test-set.
	}
	\label{fig_smmnet}
\end{figure*}

Following the previous taxonomy, our proposal would belong to the category ``\ac{dl} \acl{ds} applied to \ac{dl} \acl{af}'' .
Unlike previous feature learning approaches (that rely on a Triplet Loss paradigm), we utilize a more straightforward paradigm (illustrated in Figure~1) which is a succession of two steps, each with its own objective. 
The two objectives are jointly optimized.

In the \textbf{first step}, we learn the parameters $\theta$ of an encoder $f^{\theta}$ such that when applied to the sequence of inputs $\{\X_i\}_{i \in \{1 \ldots T\} }$ that represent a given track (where $T$ is the length of temporal sequence), the encoded features allows the estimation of a \ac{ssm}, $\hS$, which attempts to reproduce a ground-truth \ac{ssm}, $\S$.
For training $f^{\theta}$ we use an approach similar to the SSM-Net approach proposed in \cite{Peeters2022LBDSsmNet}, i.e. defining a loss which directly compare the obtained SSM $\hS$ to a ground-truth SSM $\S$.

In the \textbf{second step}, we learn a set of kernels $\K$ such that when convolved over the main diagonal of the estimated \ac{ssm} $\hS$ it allows the estimation of a novelty score $\hn$, which attempts to reproduce a ground-truth novelty score, $\n$.
This novelty score is usually obtained using a fixed checkerboard kernel~\cite{FooteIEEE2000Segmentation}. 
The resulting function is named novelty score since high values in it indicate times where the content change (it is homogeneous before and after).
It has been shown that better kernels can be used (for example using multi-scale kernels \cite{Kasier2013ICASSPMultiKernel}) and that it is possible to train such kernels $\K$ considered as the kernels of a ConvNet (for example \cite{ullrich_boundary_2014} and \cite{grill_music_2015} in the case of a (time,lag) SSM or \cite{cohen-hadria_music_2017} in the case of a (time,time) SSM, which is our case).
This is the approach we follow here.

Another proposal we make in this paper, is to consider the learning of relative features, i.e. features which are relative to the given track.

\textbf{Paper organization.}
We provide an overview of our system in part~\ref{part_proposal}, describe the inputs to our system (part~\ref{part_input}), detail the two losses (parts~\ref{part_loss_ssm} and \ref{part_loss_nov}), motivate relative feature learning (part~\ref{part_selfattention}), detail the architecture of our encoder $f^{\theta}$ (part~\ref{part_architecture}) and the training process (part~\ref{part_training}).
In part~\ref{part_evaluation}, we provide a large-scale evaluation of our proposal.
It should be noted that although we only evaluate our method for the task of segment boundary detection, it can also be used for segment labeling given the clearness of the obtained SSM.

% ---------------------------------
% ---------------------------------
% ---------------------------------
\vspace{-0.2cm}
\section{Proposal}
\label{part_proposal}

% ---------------------------------
% ---------------------------------
\subsection{Input data $\{\X_i\}$}
\label{part_input}

The inputs $\{\X_i\}$ to our system are simple patches\footnote{
We utilized patches as input (rather than frames) because we believe that homogeneity exists at the pattern level rather than the frame level.} of Log-Mel-Spectrogram.
We didn't consider beat-synchronous features as in \cite{mccallum_unsupervised_2019} given the non-reliability of beat estimation outside popular music.
Using \texttt{librosa}~\cite{mcfee2015librosa}, we first computed Mel-spectrogram with 80 mel-bands, using a 92ms window length and 23ms hop size. 
Those are then converted to log-amplitude using $\log(1\!+\!100 \cdot mel)$.
We then aggregate them (mean operator) over time to lead to a 0.1s hop size.
The final $\{\X_i\}$ are then patches of 40 successive frames (corresponding to 4s.) with a hop size of  5 frames (corresponding to 0.5s.).

%Although we said in part~\ref{sec:proposal} that our system belongs to \acf{dl} \acf{af}, we didn't actually start the feature learning from raw audio waveforms but from higher representations.

% ---------------------------------
\subsection{SSM-loss}
\label{part_loss_ssm}

Given a sequence of inputs $\{\X_i\}_{i \in \{1 \ldots T\} }$ , we apply the same encoder $f^{\theta}$ individually to each $\X_i$ to obtain the corresponding sequence of embeddings $\{\e_i\}_{i \in \{1 \ldots T\} }$.
Those are then L2-normalized.
We can then easily construct an estimated \ac{ssm}, $\hS$, using a distance/similarity/divergence $g$ between all pairs of projections: 
\begin{equation}
	\hS = g(\e_i=f^{\theta}(\X_i), \e_j=f^{\theta}(\X_j)), \;\;\; \forall i, j
\end{equation}
We use here a  ``scaled'' cosine-similarity for $g$ which, because the embeddings are L2-normalized, reduces to
\begin{equation}
	\label{eq_ssm}
	\hS=1 - \frac{1}{4} \lVert \e_i - \e_j \rVert_2^2 \;\;\; \in [0,1]
\end{equation}

It is then possible to compare $\hS$ to a ground-truth binary \ac{ssm}, $\S$, derived from annotations.
We consider this as a multi-class (a set of $T^2$ binary classifications) problem and hence minimize the sum of \ac{bce} losses.
However, given the unbalancing between the two classes in $\S$ (which contains much more 0 than 1), we used a weighting factor $\lambda$ computed as the percentage of positive values in $\S$.
The lower $\lambda$ is, the more we put emphasis on positive ($\S$=1) examples: 
\begin{equation}
	\label{eq_ssmloss}
	\L^{\theta}_{SSM}\!=\!-\!\frac{1}{T^2}\!\sum_{i,j=1}^T\! (1\!-\lambda)\!\left[\S\!\log(\!\hS\!)\!\right]\!+\!\lambda\!\left[\!(\!1\!-\!\S\!) \!\log\!(\!1\!-\!\hS\!)\!\right]
\end{equation}

Since the computation of the \ac{ssm} $\hS$ is differentiable w.r.t. to the embeddings $\{\e_i \}$, we can compute $\ddd{\L^{\theta}_{SSM}}{\theta} $:

\begin{equation}
	\ddd{\L^{\theta}_{SSM}}{\theta} 
	= \sum_{i,j=1}^T \ddd{\L^{\theta}_{SSM}}{\hS} \left( \ddd{\hS}{\e_i} \ddd{\e_i}{\theta} + \ddd{\hS}{\e_j} \ddd{\e_j}{\theta} \right)
\end{equation}
We can then use standard gradient-descent algorithms to optimize $\theta$ which will jointly optimize $f^{\theta}$ for all the $\X_i$.

Optimizing directly $\hS$ has relationship with Metric Learning / Contrastive Learning in which the $A$,$P$,$N$ are chosen based on their similarity (such as in Wang et al. \cite{Wang2021ISMIRSupervised}). 
In comparison, we consider here simultaneously all possible pairs of time as $A$,$P$,$N$.
This is actually in line with the fact that we aim at learning features relative to a track (see part~\ref{part_selfattention}) and we therefore need to consider simultaneously the interaction between all projections $\{\e_i\}$.

% ---------------------------------
\subsection{Novelty-loss}
\label{part_loss_nov}

We propose to learn the kernels $\K$ such that when convolved with the estimated \ac{ssm} $\hS$ (see 	eq.(\ref{eq_ssm})) along its main diagonal the resulting estimated novelty score $\hn$ approximate a ground-truth novelty score $\n$.
This kernel convolution can be simply implemented as an extra convolution layer (without bias) on top of the estimated \ac{ssm} $\hS$ with a sigmoid output activation.
We then define the novelty-loss as
\begin{equation}
	\label{eq_novloss}
	\L^{\theta}_{nov} = \frac{1}{T} \sum_{i=1}^T BCE(\hn, \n)
\end{equation}

% ---------------------------------
\subsection{Relative feature learning}
\label{part_selfattention}
 
In previous works dealing with feature learning for \ac{msa} it is assumed that, once trained, the network $\f$ always projects a given segment $\X_i$ in the same way whatever its surrounding context.

We advocate here that for the task of \ac{msa} the projection of $\X_i$ should  depend on its context.
The motivation for doing so is that the features that highlight the temporal structure of a music track usually depend on the track itself. 
For example, if the instrumentation or the timbre remains constant over the track, the structure may arise from variation of the harmonic content; in other cases, it will be the opposite.
Therefore, feature learning for \ac{msa} should be made relative-to-a-track.

To let each feature $\X_i$ ``know'' about surrounding times features $\{\X_1 \ldots \X_{i-1}, \X_{i+1} \ldots \X_{T} \}$ we introduce layers of \ac{sa}~\cite{all_you_need} in our encoder\footnote{Note that the use of the SSM-loss alone does not allows $f^{\theta}$ to encode relative features; this is the task of the \ac{sa}.}.

\subsection{Network architecture $f^{\theta}$}
\label{part_architecture}

The architecture of the encoder $f^{\theta}$  is given in Figure \ref{fig_smmnet}.
It is made of a succession of 5 consecutive convolution blocks followed by $N$ blocks of Transformer-Encoder.

Each convolution block is made of a 2D convolution followed by a PReLU \cite{selu} activation and a 2D max-pooling.
The kernel size $(k_f, k_t)$, the number of channels $n_c$ and  pooling size $(p_f,p_t)$) of each layer are the following: 
layer-1: $(k_f, k_t)$=(5,5) $n_c$=32 $(p_f,p_t)$=(2,2),
layer-2: (5,5) 32 (2,2),
layer-3: (5,5) 64 (2,2),
layer-4: (5,5) 64 (2,2),
layer-5: (5,2) 128 (5,2).
The output of the last convolutional blocks has dimension (1,1) with $n_c$=128 channels and is flattened to a 128-dim vector.

Each input $\X_i$ is independently projected using the convolutional blocks.
These outputs are then considered as a temporal sequence which is fed to $N$ blocks of Transformer Encoder (each made up of a \ac{sa} layer with 8 heads, skip-connection, a normalization layer and two fully-connected layers with an internal dimension of 128).
The outputs are then passed to a tanh and L2-normalized.
They form a sequence of embeddings $\{\e_i\}_{i \in \{1 \ldots T\} }$ with $\e_i \in R^{128}$ which are used to compute $\hS$.

The size of the kernels $\K$ is fixed to (41,41) which roughly corresponds to 20s.
The kernels $\K$ are either initialized randomly or initialized with checkerboard kernels similar to the ones of \cite{FooteIEEE2000Segmentation}. 
In this case, checkerboard kernels have the same size (41,41) but are damped with Gaussian function with different $\sigma$ (randomly chosen in the range $[3s,5s]$).
We used 3 different kernels $\K$ which are then combined using (1x1) convolution.
The diagonal of the resulting feature-map then goes to a sigmoid activation and is considered as the estimated novelty $\hn$.

Our architecture remains lightweight with a number of parameters ranging from 268K to 567K depending on the number of Transformer Encoder blocks (from $N$=0 to 3).

\subsection{Training.}
\label{part_training}

We train our network by minimizing jointly the two losses defined by eq. (\ref{eq_ssmloss}) and eq. (\ref{eq_novloss}):
\begin{equation}
	\label{eq_jointloss}
	\L^{\theta} = \alpha \L^{\theta}_{SSM} + (1-\alpha) \L^{\theta}_{nov}
\end{equation}

We used the ADAM optimizer with a learning rate of 0.001, used early-stopping monitoring $\L^{\theta}$ on the validation data with a patience of 50 and a maximum of 500 epochs.

Considering that we need the whole sequence of embeddings $\{ \e_i\}$ of a track to compute $\hS$ and get the gradients $\ddd{\L^{\theta}}{\theta}$, the mini-batch-size $m$ is here defined as the number of tracks.
We used a value of $m$=10 tracks.

% ---------------------------------
\subsubsection{Generating ground-truth for training}
\label{part_gtssm}

\textbf{Ground-truth SSM $\S$.}
The ground-truth \ac{ssm}, $\S$, is constructed using annotated segments (start and end time) and their associated labels.
We rely on the homogeneity assumption, i.e. we suppose that all times $t_i$ that fall within a segment are identical since they share the same label.
If we denote by $\text{seg}(t_i)$ the segment $t_i$ belongs to and by $\text{label}(\text{seg}(t_i))$ its label, we assign the value   $\S\!=\!1$ if  $\text{label}(\text{seg}(t_i)) = \text{label}(\text{seg}(t_j))$ and  $0$ otherwise.
Note that we could relax this identity constraint to allow representing similarity between labels (for example using an edit distance between labels).
This is for example important for RWC-POP dataset, where labels denotes some proximities (\texttt{verse A} and \texttt{verse B}) but are here considered as different.
Also, it could be important to consider the case of non-homogeneity of the repetitions and create a ground-truth $\S$ made of ``sub-diagonals'' rather than ``blocks''.

\textbf{Ground-truth novelty score $\n$.}
The ground-truth novelty score, $\n$, is also constructed using the annotated segments (start and end time). We set $\n$ to 1 when segment changes at time $i$, 0 otherwise.
As proposed by \cite{schluter_improved_2014} we smooth over time the boundary annotations by applying a low-pass filter with a triangular-shape $\{0.25, 0.5, 1, 0.5, 0.25\}$.

% ---------------------------------
% ---------------------------------
% ---------------------------------
\section{Evaluation}
\label{part_evaluation}

We assess here the performance of our proposal using various test sets, compare it to previously published results, conduct an ablation study, and illustrate its results.

% ---------------------------------
\subsection{Datasets}
\label{part_dataset}

For training we used a subset of 693 tracks from the \textbf{Harmonix} dataset~\cite{nieto_harmonix_2019}\footnote{Given the non-accessibility of Harmonix audio, those have been downloaded from \texttt{YouTube} and re-annotation has been necessary because of non-synchronicity of the original annotations.} and
the 298 tracks of the \textbf{Isophonics} dataset~\cite{Mauch2009ISMIRAnnotation}.
For testing we used 
%\vspace{-0.2cm}
\begin{itemize}
	\itemindent=-15pt
	\itemsep=-3pt
\item \textbf{RWC-Pop-AIST} the 100 tracks of the RWC-Pop \cite{goto_development_2004} with AIST annotations \cite{Goto2006ISMIRAIST} and the following three subsets of the SALAMI~\cite{smith_design_2011} dataset:
	\item \textbf{SA-Pop} is the subset of SALAMI tracks with CLASS equal to Popular, 
	\item \textbf{SA-IA} is the subset of SALAMI tracks with SOURCE equal to IA (Internet Archive),
	\item \textbf{SA-Two} is the subset of SALAMI tracks with at least two annotations.
\end{itemize}
For each SALAMI subset we considered the two annotations (An1, An2) and the two levels of flat annotations (Upper, Lower); those correspond to the files \verb+textfile{1,2}_{upper,lowercase}.txt+.

\begin{table}[]
	{\small
	\begin{tabular}{l|l|ll|ll}
		Datasets & T & S & L & S & L \\
		\hline
		\hline
		Harmonix & 693 & 13 & 17.15 &  &  \\
		Isophonics & 298 & 11 & 15.98 &  &  \\
		\hline
		RWC-Pop-AIST & 100 & 17 & 14.31 &  &  \\
		\hline
		\hline
		&  & Upper &  & Lower &  \\
		SA-Pop (An1) & 276 & 12 & 15.49 & 30 & 5.73 \\
		SA-Pop (An2) & 175 & 12 & 14.64 & 31 & 5.67 \\
		\hline
		SA-IA (An1) & 444 & 14 & 18.32 & 50 & 4.43 \\
		SA-IA (An2) & 244 & 12.5 & 18.67 & 37 & 7.00 \\
		\hline
		SA-Two (An1) & 882 & 11 & 18.25 & 30 & 6.89 \\
		SA-Two (An2) & 882 & 11 & 17.76 & 31 & 6.39
	\end{tabular}
	}
	\label{table_dataset}
	\caption{Description of the datasets used in our evaluation: number of tracks $T$, median value of the number of segments per track $S$ ,  median value of segment duration $L$ in seconds (note that \cite{Buisson2022IsmirStructure} indicate $L$ in number of beats).}
\end{table}

In Table~1 we describe these datasets.
According to the values of $L$ our training-sets better match the Upper annotations than the Lower ones of SALAMI.
Also, the size of our kernels $\K$ (roughly 20s., see part~\ref{part_architecture}) assumes homogeneous segments of  roughly 10s. and are therefore closest to the $L$ of Upper annotations.

\begin{table*}[t]
	\centering
{\footnotesize
	\begin{tabular}{l|ll|ll|ll|ll|l}
		& \multicolumn{2}{c|}{RWC-Pop-AIST} & \multicolumn{2}{c|}{SA-Pop} & \multicolumn{2}{c|}{SA-IA} & \multicolumn{2}{c|}{SA-Two} &  \\
		& HR.5F & HR3F & HR.5F & HR3F & HR.5F & HR3F & HR.5F & HR3F & Annotation \\
\hline
\hline
		Grill \cite{grill_music_2015,grill_structural_2015}  {\scriptsize GS1}    & .506 & \textbf{.715} & - & - & - & - & .541 & .623 &  Up./An-*\\
		McCallum \cite{mccallum_unsupervised_2019} {\scriptsize Unsynch.} & - & - & - & - & - & .497 & - & - &  \\
		\hspace{1.75cm} {\scriptsize Beat-synch.} & - & - & - & - & - &\textbf{ .535} & - & - &  \\
		Salamon  \cite{Salamon2021ISMIRDeepEmbedding} {\scriptsize $\text{DEF}_{0\!.5\!,0\!.\!5}$/$*_{\mu\!^H,\!\gamma\!^H}$}& - & - & - & - & - & - & .337 & .563 & Up./An-* \\
		Wang \cite{Wang2021ISMIRSupervised} {\scriptsize scluster/D/eu/mul} & .438 & .653 & .447 & .623 & - & - & .356 &  .553 & Up./An-* \\
		Buisson \cite{Buisson2022IsmirStructure} {\scriptsize $\text{HE}_0$/$\text{HE}_1$} & - & .681 & - & - & - & - & - & \textbf{.597 / .595} & Up./An-1/2 \\
		&  &  & - & - & - & - & - &\textbf{.611 / .600} & Low./An-1/2 \\
\hline
\hline
		\textbf{Ours} (best conf.)  & .399 & \textbf{.713} & .298 / .295& \textbf{.631/ .624} & .250 / .261 & \textbf{.520 / .511} & .231 / .237 & .521 / .530 & Up./An-1/2 \\
		&  &  & .296 / .318 & .570 / .610 & .302 / .336 & .547 / .612 & .287 / .287 & .589 / .589 & Low./An-1/2 \\
\hline
\hline
		\textbf{Ablation study $N$} &  &  &  &  &  &  &  &  &  \\
		\textbf{N=3/$\alpha$=0.5/K:train-Init:chck} &  & \textbf{.713} &  & .532 &  & .472 &  & .448 & Up./An-1 \\
		N=2/$\alpha$=0.5/K:train-Init:chck   &  & .701 &  & .535 &  & .474 &  & .449 & Up./An-1 \\
		N=1/$\alpha$=0.5/K:train-Init:chck   &  & .677 &  & \textbf{.631} &  & \textbf{.520} &  & \textbf{.521} & Up./An1 \\
		N=0/$\alpha$=0.5/K:train-Init:chck   &  & .696 &  & .535 &  & .459 &  & .443 & Up./An-1 \\
\hline
		\textbf{Ablation study $\alpha$} &  &  &  &  &  &  &  &  &  \\
		N=3/$\alpha$=1/K:train-Init:chck  &  & .154 &  & .121 &  & .102 &  & .111 & Up./An-1 \\
		N=3/$\alpha$=0/K:train-Init:chck  &  & .007 &  & .120 &  & .026 &  & .095 & Up./An-1 \\
\hline
		\textbf{Ablation study $\K$} &  &  &  &  &  &  &  &  &  \\
		N=3/$\alpha$=0.5/K:train-Init:randn    &  & \textbf{.713} &  & .543 &  & .470 &  & .457 & Up./An-1 \\
		N=1/$\alpha$=0.5/K:train-Init:randn    &  & .709 &  & .547 &  & .470 &  & .457 & Up./An-1 \\
		N=3/$\alpha$=0.5/K:fix-Init:chck  &  & .330 &  & .250 &  & .199 &  & .196 & Up./An-1
	\end{tabular}
}
	\caption{Results of segment boundary detection using various test-sets and configurations}
	\label{table_results}
\end{table*}

% ---------------------------------
\subsection{Segment detection from novelty score}

To get the estimated segment boundaries from the estimated novelty score $\hn$ we used a simple peak-to-mean ratio algorithm similar to \cite{mccallum_unsupervised_2019}.
Using the same notations as \cite{mccallum_unsupervised_2019}~eq.~(5), we compute the mean  with a  window of duration $2T$=20s, and then detect local peaks with a threshold $\tau$=1.35 and a minimum inter-distance of 7s.

% ---------------------------------
\subsection{Performance metrics}

We evaluate the performance of segment boundary detection using the common Hit-Rate metrics using precision-windows of 3s and 0.5s.
We only display here the Hit-Rate F-measures denoted by \HRt$\;$ and \HRd.
We used \verb+mir_eval+ ~\cite{Raffel2014IsmirMirEval} with \verb+mir_eval.segment.detection+ ignoring track start and end annotations (\verb+Trim=True+).
We point out that without ``trimming'' (the start and end time) we would gain +3\% on average (from .713 to .743 for RWC-Pop).

%\vspace{-0.2cm}
% ---------------------------------
\subsection{Comparison with previous works}

In the following we will compare our results with the ones previously published by 
Grill and Schlüter in \cite{grill_music_2015, grill_structural_2015},
McCallum et al. in \cite{mccallum_unsupervised_2019}, 
Salamon et al. in \cite{Salamon2021ISMIRDeepEmbedding}, 
Wang et al. in \cite{Wang2021ISMIRSupervised} and 
Buisson et al. in \cite{Buisson2022IsmirStructure}.
We first check if their test-sets match ours.

For SA-Pop, Wang~\cite{Wang2021ISMIRSupervised} used \textit{``a subset with 445 annotated songs (from 274 unique songs) in the ``popular'' genre''}. 
This roughly matches our SA-Pop (An1)+(An2) which provides 276+175=451 annotations. 
They used the Upper-case annotations (personal communication). 

For SA-IA, McCallum~\cite{mccallum_unsupervised_2019} used \textit{``the internet archive portion of the SALAMI dataset (SALAMI-IA) consisting of 375 hand annotated recordings''}. 
This is much lower than our SA-IA (An1)+(An2) which provides 444+244=688 annotations.
Moreover, it is not clear whether they used the Upper, Lower or Functional annotations.

Finally, for SA-Two, Salamon~\cite{Salamon2021ISMIRDeepEmbedding} Table 3 used the Upper-case annotations of tracks with at least 2 annotations (884 tracks); 
Wang et al.~\cite{Wang2021ISMIRSupervised} \textit{``we treat each annotation of a song separately, yielding 2243 annotated songs in total''} and 
Buisson et al.~\cite{Buisson2022IsmirStructure} used the Upper and Lower-case annotations of tracks with at least 2 annotations (884 tracks). 
This roughly corresponds to our SA-Pop (An1)+(An2) which has 882 tracks.

% ---------------------------------
% ---------------------------------
\subsection{Results and discussions} 
\label{part_results}

Results are given in Table~\ref{table_results}.
The upper part shows previously published results, although not all systems were evaluated on all test sets. 
%The sign '-' indicates that the results are not available for this dataset.
%
The middle part shows the results achieved with the best configuration of our system.

For \textbf{RWC-Pop-AIST}, we obtained a HR3F= .713\footnote{The Precision and Recall at 3seconds are P3F=.735, R3F=0.715} which is comparable to those of Grill and Schlüter (.715). However, for HR.5F our results are below (.399 < .506). 
{This can be explained by the fact that the hop-size of our features$\{\X_i\}$  was chosen large (0.5s) and does not allow to have a precise boundary estimation.
We have chosen a large hop size to reduce the size of $\hS$ (hence the computation time and memory footprints); it also allows to keep the size of the $\K$ manageable.
Because of this, all our results with HR.5F are actually low. 
Therefore, we only discuss the results for HR3F in the following.}

For \textbf{SA-Pop}, we obtained a HR3F of .631/.624\footnote{P3F=.581, R3F=0.760/ P3F=.566, R3F=0.771 $\rightarrow$ over-segmentation} for the two Upper annotations (Up./An-1/2) which is slightly above  those of Wang et al. (.623).
For the two lower annotations (Low./An-1/2) we get  a HR3F of   .570/.610\footnote{P3F=.860, R3F=0.468/ P3F=.877, R3F=0.497 $\rightarrow$ under-segment.}. 
Wang et al. does not provide these results.

For \textbf{SA-IA}, we obtained a HR3F of .520/.511\footnote{P3F=.435, R3F=0.718/ P3F=.411, R3F=0.751 $\rightarrow$ over-segment.} for the two Upper annotations and .547/.612\footnote{P3F=.811, R3F=0.451/ P3F=.756, R3F=0.546 $\rightarrow$ under-segment.} for the two Lower annotations. 
This has to be compared to the .497 (unsynchronized) and .535 (beat-synchronized) obtained by McCallum et al., but as explained, it is not clear whether they used Upper, Lower or Functional annotations.

For \textbf{SA-Two}, we obtained a HR3F of .521/.530\footnote{P3F=.433, R3F=0.749/ P3F=.442, R3F=0.754$\rightarrow$ over-segment.}   for the two Upper annotations. 
This is slightly lower than the results of Wang et al. (.553), Salamon et al. (.563), Buisson et al. (.597) and largely below the ones of Grill and Schlüter (.623).  
For the Lower annotations, we obtained a HR3F of .589/.589\footnote{P3F=.768, R3F=0.523/ P3F=.768, R3F=0.523$\rightarrow$ under-segment.}  which is slightly below the ones of Buisson et al. (.611). 
It should be noted however that in our work we didn't used any data from SALAMI, neither for training or validation (such as early stopping).

For SA-IA and SA-two, our results are higher for the Lower annotations than the Upper ones.
This is surprising since according to Table~1 the characteristics ($L$ value) of our training sets are closer to the Upper case.
Also (see  footnotes 8--14), our algorithm tends to over-segment when considering the Upper annotation and under-segment when considering the Lower ones. 
Our kernel size is actually between the $L$ values of the Upper and Lower annotations.

% ---------------------------------
% ---------------------------------
\subsection{Ablation study} 
\label{part_ablation}

In the lower part of Table~\ref{table_results} we perform an ablation study of our system.
For the SA-\{Pop,IA,Two\} test-sets, we only perform the study using the Upper/An1 annotations 

We first check the optimal number $N \in \{0,1,2,3\}$ of layers of Transformer Encoder. 
We see that for all test-sets the use of Transformer Encoder ($N>0$) is beneficial. 
For RWC-Pop-AIST, the optimal number is $N$=3 while for all three SA-\{Pop,IA,Two\} test-sets it is always $N$=1.

We then check whether jointly optimizing the two losses $L^{\theta}_{SSM}$ and $\L^{\theta}_{nov}$ of eq. (\ref{eq_jointloss}) is necessary. 
We considered three cases: $\alpha$=1 (only optimizing $L^{\theta}_{SSM}$), $\alpha$=0.5 (optimizing both), $\alpha$=0 (only optimizing $\L^{\theta}_{nov}$). 
For all test-sets, we see that optimizing jointly the two losses is highly beneficial: for example, for RWC-Pop-AIST, HR3F=.713 with $\alpha$=0.5, .154 with $\alpha$=1 and .007 for $\alpha$=0.

Finally, we check various configurations of the kernels $\K$.
$\K$ is either 
[K:train-Init:chck]: trained starting from checkerboard kernels initialisation, [K:train-Init:randn]: trained starting from random  initialisations,
[K:fix-Init:chck]: fixed (not trained) to checkerboard kernels (we still train the 1x1 convolution).
We see that for all test-sets it is beneficial to train $\K$ (the worst results are obtained with [K:fix-Init:chck]).
For RWC-Pop-AIST, the results are the same whether kernels are initialized randomly or with checkerboard kernels. 
For SA-\{Pop,IA,Two\} the checkerboard kernels initialization is beneficial.

% ---------------------------------
% ---------------------------------
\subsection{Examples} 
\label{part_example}

\begin{figure}[ht!]
	\centerline{\includegraphics[clip, trim=1cm 2cm 1cm 4cm,width=1.05\columnwidth]{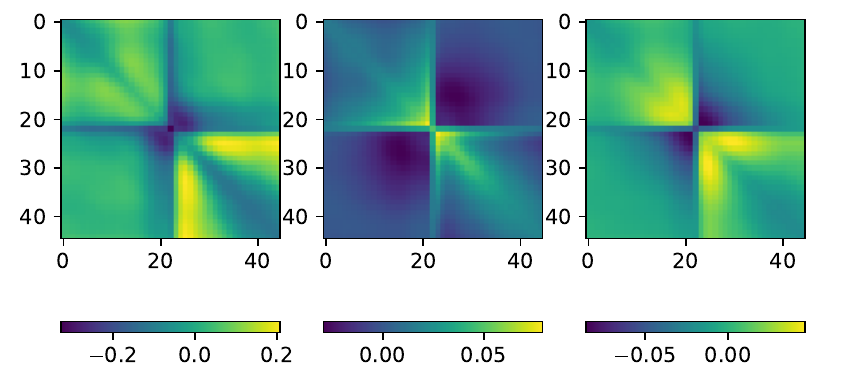}}
	\caption{The three kernels $\K$ learned using the [N=3 / $\alpha$=0.5 / K:train-Init:chck] configuration.}
	\label{fig_kernels}
\end{figure}

\begin{figure}[h!]
	\centerline{\includegraphics[clip, trim=3cm 1cm 1cm 1cm,width=.95\columnwidth]{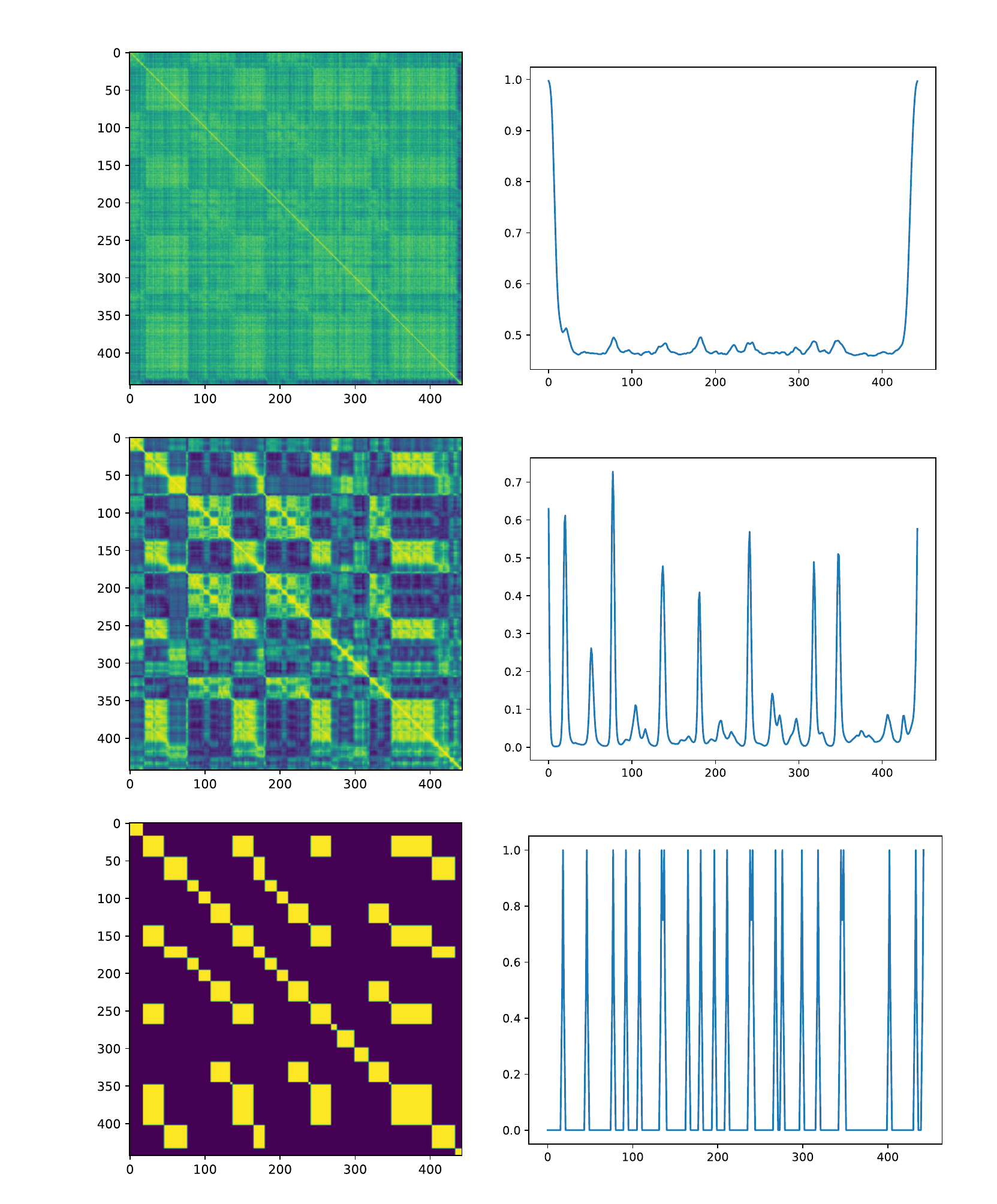}}
	\caption{[Top] $\hS$ and $\hn$ obtained with untrained system using [K:fix-Init:chck] for the kernels,
	[Middle] same with [N=3 / $\alpha$=0.5 / K:train-Init:chck], 
	[Bottom] ground-truth $\S$ and $\n$.}
	\label{fig_example}
\end{figure}

Figure~\ref{fig_kernels}  illustrates the three kernels $\K$ learned using the [N=3/$\alpha$=0.5/K:train-Init:chck] configuration.
As one can see, while the middle one looks close to the classical checkerboard kernel of Foote~\cite{FooteIEEE2000Segmentation} (but with an emphasis on the diagonal), the first seems to highlight the transition from a non-homogeneous to an homogeneous part; while the third seems a re-scaled version of the second (homogeneity at a larger scale).
Figure~\ref{fig_example} illustrates the $\hS$ and $\hn$ obtained by our system on track-01 from RWC-Pop-AIST (chosen as the first item of our test-set).
We compare the results when trained in the [N=3 / $\alpha$=0.5 / K:train-Init:chck] configuration and with the untrained system using [K:fix-Init:chck] for the kernels. 
For comparison we indicate the ground-truth $\S$ and $\n$.
In this figure, the benefits of training both $L^{\theta}_{SSM}$ and$\L^{\theta}_{nov}$ appears clearly.

\paragraph{Reproducibility.} The code and the datasets used in this work are available at:  \href{https://github.com/geoffroypeeters/ssmnet_ISMIR2023}{https://github.com/geoffroypeeters/ssmnet\_ISMIR2023}


% ---------------------------------
% ---------------------------------
% ---------------------------------
%\vspace{-0.3cm}
\section{Conclusion}

In this work, we proposed a simple approach for \acl{dl}-based \acl{msa} algorithm:  we learn an encoder $\f$ such that the resulting learned features allow to best approximate a ground-truth \ac{ssm};  we jointly learn segmentation kernels that when applied to the estimated \ac{ssm} we best approximate a ground-truth novelty score.
We also propose to learn relative features, i.e. features relative to a track, by introducing \acl{sa} layers in our encoder.
According to HR3F, our results  are either better than previous state-of-the-art (SA-Pop, SA-IA unsynchronous), similar (RWC-Pop-AIST) or worst (SA-Two).
%According to HR3F, our results are comparable to previously published ones.
Our approach has the advantage to be lightweight (around 500K parameters) and based on criteria which are semantically linked to the task of \ac{msa}.
Future works will concentrate on making our approach applicable to finer temporal resolutions, therefore allowing improving our performances at HR.5F.

\clearpage
\bibliography{smmnet}

\end{document}